# One Size Doesn't Fit All: When to Use Signature-based Pruning to Improve Template Matching for RDF graphs

Shi Qiao, Z. Meral Ozsoyoglu

Computer Science, Case Western Reserve University
{sxq18, mxo2} @case.edu
Cleveland, OH, 44106

**Abstract.** Signature-based pruning is broadly accepted as an effective way to improve query performance of graph template matching on general labeled graphs. Most techniques which utilize signature-based pruning claim its benefits on all datasets and queries. However, the effectiveness of signature-based pruning varies greatly among different RDF datasets and highly related to their dataset characteristics. We observe that the performance benefits from signature-based pruning depend not only on the size of the RDF graphs, but also on the underlying graph structure and the complexity of queries. This motivates us to propose a flexible RDF querying framework, called RDF-$\hbar$, which selectively utilizes signature-based pruning by evaluating the characteristics of RDF datasets and query templates. We demonstrate the scalability and efficiency of RDF-$\hbar$ via experimental results using both real and synthetic datasets.

## 1 Introduction

In this paper, we study the effectiveness of signature-based pruning for querying graph-structured data using graph templates focusing on graph-structured RDF data. Neighborhood signature-based pruning has been used extensively to improve performance of graph template matching (based on subgraph isomorphism) problem. Many different variations of neighborhood signature indexes have been developed [15-20]. While signature-based pruning can be beneficial for queries on RDF datasets, there are several factors to be considered. First, RDF graphs have unique node labels, and use URIs to identify millions of resources which are typically long strings, and partial(ly entered) keywords are used in specifying RDF query templates. Thus, the design of neighborhood indexes and the procedure of neighborhood containment check in query processing need to support queries with partial keywords. And, effectiveness of signature-based pruning highly depends on the characteristics of datasets and queries [7]. The underlying graph structure of RDF datasets can range from strict relational like structure to arbitrary graphs for different applications. Thus, signature-based pruning needs to be used selectively for different RDF graphs and queries. We focus on three query evaluation criteria for RDF graphs, namely, (1) *flexibility and expressiveness of*

*query templates*: query templates that are flexible to specify both keywords and the graph structure; (2) *utilization of the characteristics of dataset and query templates* for query optimization; and (3) *scalable query evaluation* in order to scale to RDF datasets with millions or more triples.

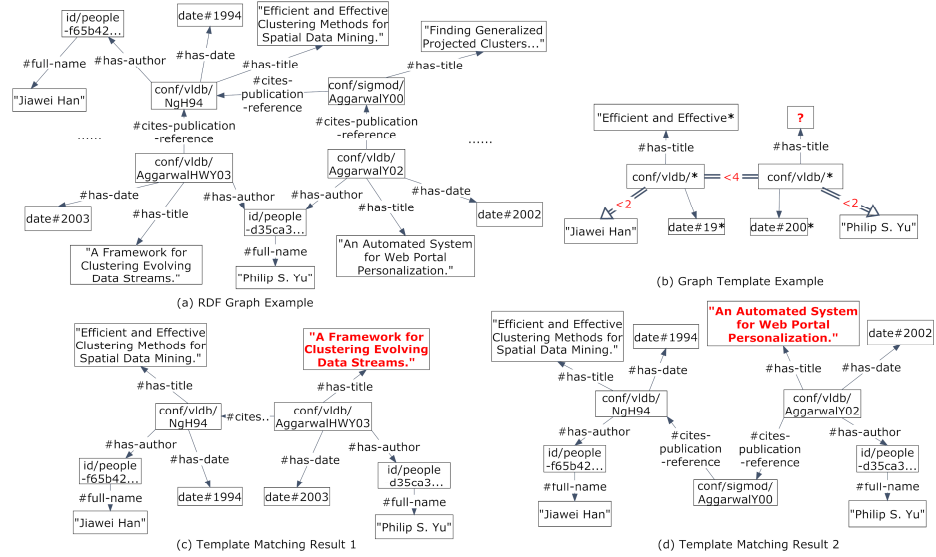

**Fig. 1.** RDF Graph and Query Template with Matching Results

### 1.1 Template Matching for RDF graphs

Here we define graph template matching for RDF graphs with a flexible query template supporting paths, distance constraints, and partial matching of keywords:

An **RDF Graph** is a directed graph $G = \{V, E, l, f\}$ where $V$ is a set of nodes representing subjects, objects or both. $E \subseteq V \times V$ is a set of directed edges representing predicates pointing from subjects to objects. $l$ is a label set for subjects, objects and predicates. $f: V/E \to l$ denotes the mapping function between nodes/edges to labels.

A **connection edge** ($\overset{E}{\Leftrightarrow}$) represents a path $\omega_{i,j}$ between two nodes $n_i$ and $n_j$, where $\omega_{i,j}$ can be one directional or bi-directional. Expression $E$ describes the distance constraints of $\omega_{i,j}$ (distance is the length of the shortest path between two nodes).

A **Query Template** is a directed graph $G_q = \{V, E\}$, where nodes (subjects or objects) are labeled by partial keywords (substrings of labels in the label set $l$ of RDF graph $G$), and edges represent predicates or connection edges.

Given an RDF graph $G = \{V, E, l, f\}$ and a query template $G_q = \{V, E\}$, **Template Matching** finds all subgraphs of $G$ satisfy both structural and label constraints in $G_q$.

An example query template is shown in Figure 1(b) on RDF graph of Figure 1(a). The query template finds the title of a paper authored by "Philip S.Yu", published in VLDB, and has a connection within 4 hops of a paper by "Jiawei Han". Using the RDF graph in Figure 1(a), two matches are shown in Figures 1(c-d). The first result is a paper directly referenced by the second paper from "Jiawei Han", and the second result is a paper which is connected to it with a 2 hop path. Connection edges are also used between Paper ID and its author name in the query template to handle Ontology nodes (People ID) between them. This example demonstrates the uses of (i) connection edges in query templates and (ii) partial keywords for querying data instead of detailed labels. Similar query templates with connection edges are also used elsewhere [26, 27], with the differences: (i) multi-attribute labels are used, but partial keyword is not supported; (ii) graph simulation is used, instead of graph isomorphism.

### 1.2 Related Work

Recently, the functionalities of SparQL [1], the standard query language of RDF data, is extended to support paths defined with regular expressions (*property paths*). Most existing RDF management systems [2-5] do not build specific indexes to handle *property paths*. Neighborhood signature index is considered as a good candidate since it can: 1) reduce the unnecessary candidates in subgraph isomorphism matching, and 2) handle connection edges (*property paths*) with short distance constraints efficiently. The idea of signature-based pruning is to check whether the labels of neighbors for a query node is contained in the labels of neighbors of its matching candidates. TALE [15] and SAPPER [16] use similar techniques where the labels of neighbors for each node are indexed by hashing into bit arrays and these bit arrays are utilized as signatures to check the neighborhood containment. Bloom filter can be utilized to generate more compact neighborhood signatures. GraphQL [17] and SPath [18], on the other hand, use neighborhood indexes by directly indexing node labels. In GraphQL, the neighbors of each node are indexed as a sequence of node labels in lexicographic order. The signature index utilized in SPath group the same labels of neighbors of each node into one entry and counts the occurrences of each label. gStore [19, 20] (i) extends the utilization of signature-based pruning to RDF databases which stores the RDF graph in the form of adjacent lists, (ii) uses the neighborhood signature as a bitstring according to the adjacent edge labels and node labels, and (iii) indexes all vertex neighborhood signatures using a special index schema, VS tree, to provide efficient query evaluation. Some recent works [21, 22] attempt to build workload sensitive indexes to query RDF data. Scalability issues of querying large RDF datasets are addressed in [23-25] by utilizing a distributed framework.

### 1.3 Our Contributions

We propose a query evaluation framework which selectively utilizes signature-based pruning depending on the characteristics of datasets and query templates. Since signature-based pruning is not effective for all datasets [7], it is important to use it when it improves the query performance. Our contributions are:
- Defining parameters that describe dataset characteristics and complexity of query template to be used in evaluating the effectiveness of signature-based pruning;

- Optimized query execution on the proposed query framework RDF-ℏ(RDF-Hybrid) which selectively uses signature-based pruning based on its effectiveness. RDF-ℏ also handles partial keywords and connection edges efficiently;
- Identifying dataset evaluation metrics that indicate levels of expected performance gains resulting from signature-based pruning in query evaluation;
- Experimental evaluations demonstrating the scalability and efficiency of RDF-ℏ, using real and synthetic datasets.

Section 2 introduces the indexes. RDF querying framework is introduced in section 3. Section 4 presents parameters that describe dataset features and query complexity which provide metrics for the effectiveness of signature-based pruning in query evaluation. Optimized query execution plan, RDF-ℏ, is then proposed. Section 5 introduces three dataset evaluation metrics which provide insights on the levels of expected performance gains resulting from signature-based pruning in query evaluation. Experimental results are shown in Section 6, and section 7 concludes.

## 2   Indexes

We utilize two indexes namely, IDMap and NI (neighborhood interval) Indexes, to support efficient evaluation of partial keywords and connection edges.

Given two vertices $n_i$ and $n_j$ in graph $G$, if there is a directed path from $n_i$ to $n_j$, $n_j$ is a ***forward neighbor*** of $n_i$ and $n_i$ is a ***backward neighbor*** of $n_j$.

Given an RDF graph G, an ***IDMap*** Index is a mapping function $H: l_i \in l(G) \rightarrow \{j | j \in Int\}$ with two properties:

1. The set of IDs for all RDF labels forms an interval of consecutive integers;
2. IDs of labels are assigned in lexicographic order.

IDMap Index basically maps RDF labels into integer IDs in lexicographic order. For partial keywords specified as prefixes of RDF labels, the look-up time is $O(logN)$, where $N$ is the total number of RDF labels, since all matching IDs form one interval of consecutive integers. Other indexes can also be built to accelerate look up time of partial keywords.

The ***NI*** index is built based on the IDMap index by grouping the labels (IDs) of neighbors of each node into ID intervals. It is designed as a table with five columns: for any node $n_i \in G$, it contains ID of $n_i$, Distance, Label ID interval, Number of indexed neighbor nodes in this entry, and neighbor node IDs. The Distance is the length of the shortest path from $n_i$ to the indexed neighbor node. The positive (negative) distance indicates that the indexed node is a forward (backward) neighbor. There are two predefined parameters for the NI index: ***maximum indexed distance*** $d_{max}$ and ***binning factor*** $m$. The neighbor nodes sharing the same distance are grouped together, ordered by their IDs, and partitioned into rows by the binning factor *m*, which limits the maximum number of indexed neighbor nodes in each index entry. Increasing the maximum indexed distance $d_{max}$ dramatically increases the space of NI index by including more indexed neighbors for each node. But, larger $d_{max}$ means more local information indexed which improves both the pruning power and the process time for connection edges with long distance constraints. NI index is designed to be most effective when partial keywords are specified as prefixes of RDF node labels. NI index can be viewed as a general form of the signatures utilized in both GraphQL [17] and SPath [18]. NI index also provides efficient evaluation for connection edges as multi-hops neighbors are indexed by using the node IDs.

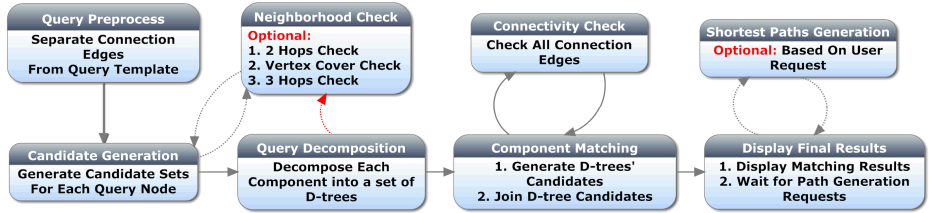

**Fig. 2.** Query Framework Structure

## 3   RDF Query Framework

The RDF query framework [Figure 2] utilizes D-tree[1] as basic querying units. The framework begins by separating connection edges from query templates, which may result in a template with several connected components. IDMap Index is then used to find matching candidates for each query node by checking the partial keywords. Neighborhood check process is then selectively chosen to prune the set of candidates for each query node. Each connected component is decomposed into a set of D-trees. All matching candidates for each decomposed D-tree are generated, and then joined together to get the matching results for each component. Connection edges are finally processed using NI index to generate final matches. Based on user requests, connection edges are instantiated by enumerating all shortest paths.

### 3.1   Neighborhood Containment Check

The neighborhood containment check for NI index is defined as ID interval check (here, we assume partial keywords are specified as prefixes of node labels).

Given a partial keyword $\wp_j$ of any query node $q_j \in V(G_q)$, the ***ID Interval***, $\mathbb{Z}_j$, is all IDs of node set $N_j$ where $\forall n_i \in N_j$, the label $l_i$ of $n_i$ is a valid match of $\wp_j$.

Given any node $n_i \in V(G)$, the ***K-Neighbor*** of $n_i$, denoted as $Neighbor_k(n_i)$, is a set of nodes where, if $k$ is positive, $\forall\, n_j \in Neighbor_k(n_i)$, there is a directed path from $n_i$ to $n_j$ with no more than $|k|$ hops; if $k$ is negative, $\forall\, n_j \in Neighbor_k(n_i)$, there is a directed path from $n_j$ to $n_i$ with no more than $|k|$ hops.

DEFINITION 3.1: Given a node $n_i \in V(G)$, and a query node $q_j \in V(G_q)$, $n_i$ passes the ***Neighborhood Check*** of $q_j$ if $\forall\, q_k \in Neighbor_k(q_j)$, ID Interval $\mathbb{Z}_k$ uniquely contains ID of any $n_g \in Neighbor_k(n_i)$, for all $|k| \leq d_{max}$.

Each partial keyword in $G_q$ is checked by IDMap index to form an interval of consecutive integers. Since query templates can contain query nodes with the same partial

---
[1] D-tree is defined as directed one level tree (height 1 tree with directed edges).

keyword, value pairs as {Distance, Count (total appearance within Distance)} for each partial keyword are maintained for each query node. The neighborhood check is performed based on partial keywords one by one, and the count of occurrences of each partial keyword is taken into consideration. The term "uniquely contains" in the definition 3.1 means that the node $n_g$ cannot be used to match more than one ID interval. If one partial keyword contains another partial keyword, Count in value pairs is updated. **Algorithm 1** shows the neighborhood check process, {Distance, Count} pairs associated with query node $q_j$ and partial keyword $\wp_i$ are denoted as $\psi_j^i$.

| Algorithm 1: Neighborhood Check $(n_i, q_j)$ |
|---|
| Input: $n_i \in V(G)$, $q_j \in V(G_q)$, ID Intervals $\mathbb{Z}^*$, {Distance, Count} $\psi_j^*$, NI Index $I_i$ |
| Output: If $n_i$ pass neighborhood check of $q_j$, return **true**; Otherwise, return **false** |
| 1    FOR all $k$ where $|k| \leq d_{max}$ |
| 2        FOR any partial keyword $\wp_k$ |
| 3            FOREACH value pair $\{d,c\}$ in $\psi_j^k$ |
| 4                Extract all entries from $I_i$ where ID interval intersect with $\mathbb{Z}_k$ and Distance $\leq d$; |
| 5                Count all IDs in $\mathbb{Z}_k$ as $c'$; |
| 6                IF $c' \leq c$, RETURN **false**; |
| 7    RETURN **true** |

Neighborhood check is optimized for partial keywords: 1) only index entries with Label ID interval intersecting with the ID interval of the partial keyword needs to be retrieved; 2) all IDs in the index entries are valid matches for the partial keyword if the ID interval of the partial keyword contains the Label ID interval.

### 3.2 Component Matching

The component matching contains 3 main steps (see **Algorithm 2**): query decomposition, D-tree candidate generation, and D-tree candidate join. Time complexity of **Algorithm 2** is proportional to $\prod_{i=1}^{K} |C_{t_i}|$ where $K$ is number of decomposed D-trees and $|C_{t_i}|$ is the number of matching candidates for each D-tree. Finding decomposition with minimum number of D-trees is likely to improve the time complexity; however it is equivalent to the vertex cover problem [23]. Here, in step 1, we use the 2-approximation algorithm to generate D-tree decomposition [8]. Basically, an edge $(q_i, q_j)$ is picked recursively from the query component, and D-trees rooted at $q_i$ and $q_j$ are added to the result. We define selectivity function $S(q_j) = \frac{deg(q_j)}{|C_{q_j}|}$ which considers query node's degree and its corresponding candidate set size as a measurement of the priority a query node to be selected as root nodes for two reasons: (i) choosing large degree nodes first is likely to yield better results since D-trees rooted at these nodes can cover more edges in the query component which may lead to a smaller $K$ value; (ii) choosing nodes with small candidate sets first is likely to yield less matching candidates for a D-tree. In step 2, NI indexes for all possible root nodes of a decomposed D-tree are checked to generate all D-tree candidate matches. Step 3 joins all D-tree candidates together to form component matches. We define the join process as $join(C_i, C_j)$, where $C_i$ and $C_j$ are candidate sets for two subgraphs of $G_c$ (either a decomposed D-tree or joined D-trees). $Join(C_i, C_j)$ combines each pair of matches from two candidate sets by evaluating the predicate: all shared query nodes of two candidate matches need to have equal matching IDs to join. In order to improve the join performance, a new join order $J_T$ for the decomposed D-trees is used as follows: 1. begin with D-tree $t_i$ with smallest candidate set and add $t_i$ to $J_T$; 2. add D-tree $t_j$ with smallest candidate set to $J_T$ which connects to already selected D-trees in $J_T$.

| Algorithm 2: Component Matching |
|---|
| Input: Query $G_c$, All Query Node Candidate Sets $C_{q_*}$, RDF Graph $G$ |
| Output: Component Candidates $C_{G_c}$ |
| 1     Decompose $G_c$ into a set of 1 level D-trees recursively: (Pick $E(q_i, q_j) \in G_c$ with largest $S(q_i) + S(q_j)$; Add D-trees $t_i$ and $t_j$ into $T_{olt}$; Remove all edges in $t_i$ and $t_j$ from $G_c$) |
| 2     For each $t_i \in T_{olt}$ <br>        Check $Neighbor(n_i)$ for each $n_i \in C_{q_i}$ to generate candidate matches $C_{t_i}$ for D-tree $t_i$ |
| 4     Join all D-tree candidates $C_{t_*}$ based on Order $J_T$ |

A similar approach is used in STWIG [23] with the differences: 1) D-trees are used as basic join units; 2) new selectivity function is defined based on the size of candidate sets; 3) the NI index is used to generate all D-tree candidates; 4) tree join order is determined by the sizes of tree candidate sets.

### 3.3 Connectivity Check

Connection edges can be inside of a query component or joining two different components. If the edge is inside of one component then the connectivity check is used to prune the component candidates. Otherwise, connectivity check is used to determine whether the two component candidates can join or not. For inside component connection edges, the number of connectivity checks is exactly the size of the component candidate set. For a connection edge between components, the number of connectivity checks depends on the product of the sizes of components' candidate sets. In the worst case, if we have a sequence of N components to be joined by connection edges, the number of connectivity checks that need to be performed can be as large as $\prod_{i=1}^{N} |C_{G_{c*}}|$. In order to improve query performance, two rules are utilized to determine the order to process connection edges: 1) inter-component connection edges are processed before intra-component connection edges; 2) intra-component connection edges are processed in the order of the smallest product of candidate sets first.

| Algorithm 3: Connectivity check $(n_i, n_j, d_c)$ |
|---|
| Input: $n_i \in V(G)$, $n_j \in V(G)$, Maximum Path Distance $d_c$, NI Index $I_i$ and $I_j$ |
| Output: If there is a path from $n_i$ to $n_j$ with distance no greater than $d_c$, return **true**; Otherwise, return **false** |

```
1    Extract entries from $I_i$ where 0 < Distance ≤ $Ceil(\frac{d_c}{2})$ and
     combine all IDlists as $\psi_i$
2    Extract entries from $I_j$ where $Ceil(\frac{d_c}{2}) - d_c$ ≤ Distance < 0
     and combine all IDlists as $\psi_j$
3    Intersect $\psi_i$ and $\psi_j$ as $\psi_{ij}$
4    If ($\psi_{ij}$ != $\phi$)
5            Return true
6    Else Return false
```

As shown in **Algorithm 3**, processing connection edges is based on the NI index. If the maximum indexed distance $d_{max}$ of NI index is smaller than $Ceil(\frac{d_c}{2})$ ($d_c$ is the distance constraint of the connection edge), we need to combine more index entries of $n_i$'s neighbor nodes together with $I_i$ in order to get more hops of neighborhood information for $n_i$. Based on our observations, long paths are rare for RDF graphs, and this justifies our assumption of limiting the maximum indexed hops to 3 for NI index.

## 4 Signature-based Pruning Evaluation

Signature-based pruning reduces the number of intermediate candidates and joins. This advantage comes with extra space cost, and the time overhead of performing the neighborhood check process. Space required for NI index can be reduced by building multi-hop NI indexes, where large hops of neighbors are indexed not for all nodes, but only for a special subset of nodes, e.g., the ones in the vertex cover. Using the set of vertices in the vertex cover to build multi-hop indexes is a compromise between the space cost and the benefit in performance of neighborhood check with the additional cost of handling two types of vertices. Instead of finding the minimal vertex cover, we use 2-approximation vertex cover, as in [8]. The time overhead to perform neighborhood check also influences query performance. Since the effectiveness of signature-based pruning varies by the datasets and queries used, evaluating its benefit and cost to use it selectively is important.

### 4.1 Dataset Features

We utilize two features, ***Predicate Selectivity*** and ***Literal Selectivity,*** describing properties of a RDF dataset in query evaluation. Both ***Predicate Selectivity*** and ***Literal Selectivity*** are pre-computed statistics over the dataset used to estimate the potential pruning power of the neighborhood structure for a given query template. We first give some basic definitions here: For an RDF graph $G$ $(V, E)$, any node $n_i \in V$ associated with URI reference label is denoted as ***Resource Node*** (blank nodes are treated as resource nodes). Other nodes associated with literal labels are denoted as ***Literal Nodes***. Any edge $e_i \in E$ between two resource nodes is treated as ***Relationship Edge***. Other edges (from resource node to literal node) are treated as ***Attribute Edge***s. (Literal nodes has no outgoing edges in the RDF graph).

***Predicate Selectivity*** for a predicate $p$ in a RDF graph $G$ is defined as $s(p) = \frac{|p|}{|E|}$, where $|p|$ is the number of occurrences of predicate $p$ in $G$, and $|E|$ is the total number of edges in $G$. ***Predicate Selectivity*** helps to determine the pruning power of a given predicate. Lower $s(p)$ indicates rare occurrence of a predicate which also indicates high pruning ability to reduce the matching candidates by checking predicate $p$.

Since literal labels are frequently used as keywords to query RDF graphs, similarity of these labels in a given RDF graph is another factor influencing query performance. Given a length $n$ partial keyword associated with predicate $p_a$, the average number of literal labels matching this partial keyword in a given RDF dataset can be used to estimate literal similarity. However, enumerating all possible $n$-grams for literal labels of a large RDF graph is infeasible, and most of these $n$-grams are not meaningful and won't be utilized as query keywords. Here, we use all prefix $n$-grams ($n$-gram begins with the first character in a literal label) as candidates for partial keywords. Let $T_{n,p_a}$ be the set of all prefix $n$-grams of literals associated with attribute predicate $p_a$, and $m_{n,p_a}$ be the average number of literals associated with $p_a$ matching a prefix $n$-gram in $T_{n,p_a}$. ***Literal Selectivity*** of an attribute predicate $p_a$ is defined as $f_{n,p_a} = \frac{m_{n,p_a}}{|l(p_a)|}$, where $|l(p_a)|$ is the no. of unique labels for predicate $p_a$.

For large RDF datasets or RDF datasets with frequent minor updates, ***Predicate Selectivity*** and ***Literal Selectivity*** can be pre-computed by sampling sets of triples from the RDF datasets and using the average Predicate Selectivity and Literal Selectivity from the sampled triples as an estimation.

### 4.2 Query Template Complexity

Signature-based pruning tends to be effective if a query template is relatively complex and neighborhood structures associated with some query nodes have high selectivity. Two parameters, ***Iteration Threshold*** and ***Join Threshold*** are defined to identify query templates that are relatively complex with respect to the number of intermediate matching candidates, and joins in query execution respectively. The number of candidates of the root node of a decomposed D-tree determines the number of iterations to generate its candidates (see **Algorithm 2**). Since higher number of iterations often indicates more D-tree candidates will be generated, it can be used as an estimation of the number of candidates for a decomposed D-tree. We also use the product of the sizes of candidate sets of all root nodes to estimate the number of intermediate joins. If the number of candidate generating iterations for any decomposed D-tree exceeds ***Iteration Threshold*** $\tau_1$ or the estimated number of intermediate tree joins exceeds ***Join Threshold*** $\tau_2$, the query template is considered as complex and has potential room for signature-based pruning to improve query performance. In order to estimate the potential pruning power of the neighborhood structure for the query template, ***Neighborhood Selectivity*** is defined.

DEFINITION 4.3: Given a query template $G_q$ for a RDF graph $G$, the ***Neighbourhood Selectivity*** $N_{q_i}$ of any query node $q_i \in G_q$ is defined as:

$$N_{q_i} = |\sum_{p_r \in Neighbor_k(q_i)} \ln(s(p_r)) + \sum_{p_a \in Neighbor_k(q_i)} \ln(s(p_a) \times f_{n,p_a})|$$

where $p_r$ is any relationship predicate, $p_a$ is any attribute predicate, $k$ is the number of hops of neighborhood check process, $n$ is the length of partial keywords associated with edge $p_a$, $s(p_r)$ and $s(p_a)$ are predicate selectivity for $p_r$ and $p_a$, and $f_{n,p_a}$ is literal selectivity.

The *Neighborhood Selectivity* of a given query node $q_i$ estimates the probability that all predicates appear in its K-Neighbors co-incidentally associated with a node in the original RDF graph $G$. It also takes partial keywords associated with attribute predicates into consideration. As this estimated probability often turns to be very small ($0 < N_{q_i} \ll 1$), we take the absolute value of its natural logarithm as the measurement to simplify its computation and comparison. Larger *Neighborhood Selectivity* indicates rareness of the neighborhood structure associated with $q_i$ appearing in $G$, implying potentially high pruning rate to reduce the no. of matching candidates by checking the neighborhood structure.

### 4.3 RDF-$\hbar$

RDF-$\hbar$ algorithm optimizes the query execution of a query template by selectively using signature-based pruning based on two conditions: (i) whether, for this query template, the number of intermediate candidates and joins can be reduced; and (ii) whether the neighborhood structure of a query template has high selectivity to provide effective pruning. If both conditions are true, the signature-based pruning is utilized in processing the query template. After query decomposition (Figure 2), the number of candidate generating iterations for each D-tree and the estimated number of intermediate tree joins are checked with **Iteration Threshold** $\tau_1$ and **Join Threshold** $\tau_2$. If any number exceeds the threshold, then neighborhood check is considered. The next step is to compute the *Neighborhood Selectivity* for all query nodes to check if there is any query node with *Neighborhood Selectivity* reaching the minimal **Selectivity Threshold** $\tau_3$. If so, neighborhood check is utilized. The choices of thresholds $\tau_1, \tau_2$ and $\tau_3$ influence the query performance of the RDF-$\hbar$; and their values can be tuned by using a sample set of query templates on the given dataset. Details of the tuning process is available in [28].

## 5 Dataset Evaluation

The performance gains of signature-based pruning differ from dataset to dataset significantly [7]. We utilize three dataset evaluation metrics to provide insights on identifying the expected level of performance gains from signature-based pruning as *dataset coherence*, *relationship specialty* and *literal diversity*.

**Dataset Coherence** [9] is utilized to measure the structuredness of a RDF graph. Higher Dataset Coherence values indicate a RDF dataset more relational like while lower Dataset Coherence values indicate complex graph structure. The coverage metric which measures how uniformly predicates are distributed among the same type of resources in a RDF dataset is utilized in computing Dataset Coherence in a weighted formula. Dataset Coherence values is an important indicator whether the neighborhood structure associated with resources of the same type is diverse [9]. Higher Dataset Coherence values often mean similar neighbor patterns are distributed in the RDF graph and signature-based pruning is not preferred.

**Relationship specialty** measures whether there exist special entities which engage in some relationships significantly more than others [14], such as a person who publishes hundreds of papers. High relationship specialty often indicates diverse relationship patterns in a RDF graph. To compute the relationship specialty, the distribution $d_i$ which indicates the number of occurrences of a relationship $r_i$ associated with each node is analyzed. The Pearson's kurtosis [6] value of $d_i$ is utilized to measure relationship specialty of $r_i$. Relationship specialty of a dataset is defined in the form of a weighted sum of relationship specialty values of all the relationship predicates in the dataset as
RS(D) = $\sum_{r_i \in Pr}(\frac{|r_i|}{\sum_{r_j \in Pr}|r_j|} \times \mathcal{K}_{r_i})$, where $|r_i|$ is the number of occurrences of relation predicate $r_i$, $\mathcal{K}_{r_i}$ is the Pearson's kurtosis of $r_i$ and $Pr$ is the set of all relationship predicates in D.

**Literal diversity** measures the similarity among literal labels of a RDF dataset [14]. A domain specific RDF dataset often utilizes a much smaller set of frequent terms for its literal labels compared with a dataset including resources from all fields, e.g., a biology dataset vs. Wikipedia dataset. Literal diversity is measured by the number of unique "words" $w_\mathcal{M}$ contained by the literals from random sampled attribute edges with size $\mathcal{M}$ (in Table 1, $\mathcal{M}$=100,000 as in [14]).

**Dataset Analysis**

Dataset coherence, relationship specialty and literal diversity provide hints on the effectiveness of neighborhood check for different RDF datasets. High dataset coherence and low relationship specialty often indicates similar neighborhood structures are uniformly distributed among resources. Low label diversity which indicates small number of unique words in a RDF graph reduces the potential pruning rate of checking partial keywords. Based on experiments, we found datasets with high coherence, low relationship specialty and low label diversity are not likely to benefit much from the signature-based pruning due to the uniform graph structure and literals associated with all resources.

## 6      Experimental Analysis

Implementation used Visual C# and SQL Server. All experiments were performed on a 2.93GHZ Intel(R) Xeon machine with 48GB ram running Windows Server 2008. STWIG [23] is revised as STWIG+ to support partial keywords and directed graph. An improved version of SPath (named SPath(NI2)) utilizing 2 hop NI index instead to optimize neighborhood check for partial keywords, and supporting directed graph is implemented. GraphQL which uses NI indexes and not shown in experiments, performs similarly as SPath(NI2). Different neighborhood checks are employed by RDF-$\hbar$, namely, $\hbar$-2Hops (2 hop neighborhood check), $\hbar$-3Hops (3 hop neighborhood check) and $\hbar$-VC (neighborhood check based on vertex cover NI index).

Four RDF datasets, LUBM [11] (Lehigh University Benchmark), SP2B [12] (SPARQL Performance Benchmark), IMDB (Internet Movie Database, as in [10]) and DBLP [13] (Computer Science Bibliography Database) are utilized. The RDF dataset characteristics of these datasets are shown in Table 1. As seen, the *relationship specialty* of different datasets distinguish real datasets and synthetic datasets effectively. And the LUBM dataset has a high *coherence*, low *relationship specialty* and low *label diversity* compared with other datasets. Thus, we expect low benefit from signature-based pruning for the LUBM dataset, also confirmed in experiments.

Table 1. RDF Dataset Characteristics

| Dataset | Coherence | Specialty | Label Uniformity (5-gram prefixes, $\mathcal{M}=100000$) |
|---------|-----------|-----------|--------------------------------------------------------|
| IMDB    | 0.172     | 905.01    | 18335                                                  |
| DBLP-1  | 0.843     | 19905.5   | 49922                                                  |
| LUBM    | 0.938     | 1.07      | 15                                                     |
| SP2B    | 0.796     | 1.17      | 97796                                                  |

Queries are generated by first randomly selecting a subgraph from the original RDF graph, and then generalizing labels of nodes to get partial keywords. Subgraph selection begins with a small subgraph (a node, an edge or a path) and recursively selects a random edge adjacent to the result subgraph until query template size requirement is reached. Once a subgraph is selected, the partial keywords are generalized from the RDF labels and assigned to query nodes. For URI labels of resource nodes, partial keywords are generated by removing the long IDs. For literal labels, partial keywords are generated by recursively removing the last character until it matches 1 to 200 labels in the RDF graph (if more than 1 partial keyword can be generated, we randomly choose one of them). The iteration threshold $\tau_1$, join threshold $\tau_2$ and neighborhood selectivity threshold $\tau_3$ are tuned by using a set of sampled queries [28].

### 6.1 Space Comparison for Different Indexes

The average space required for NI index is $O(N \left(\frac{\mu}{2}\right)^{d_{max}} / m)$ where $d_{max}$ is the maximum indexed distance, $m$ is the binning factor, $N$ is the number of nodes in the RDF graph and $\mu$ is the average node degree. Figure 3 shows information on different indexes for different RDF datasets in percentage of the original dataset. For all different NI indexes, the binning factor m is set as 5. IDMap index is relatively small compared with NI indexes. Four different NI indexes are compared here, and are utilized by different querying techniques (1 hop index for "STWIG+", 2 hop index for "SPath(NI2)" and "ℏ-2Hops", 3 hop index for "ℏ-3Hops" and vertex cover index for "ℏ-VC"). Clearly, the space needed for NI indexes increases as $d_{max}$ increases. The vertex cover index uses 2 hop neighbors for nodes in the vertex cover, and 1 hop neighbors for other nodes. Thus, the space requirement for the vertex cover index is between 1 hop index and 2 hop index. The average node degree of the RDF graph also impacts index sizes.

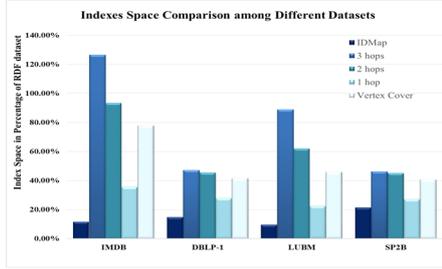
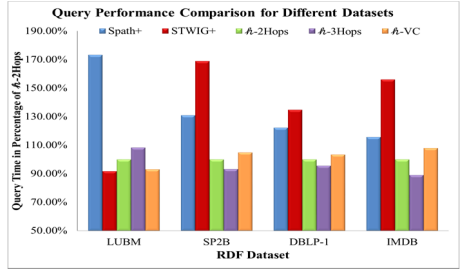

Fig. 3. Indexes Space Comparison    Fig. 4. Query Comparison (Datasets)

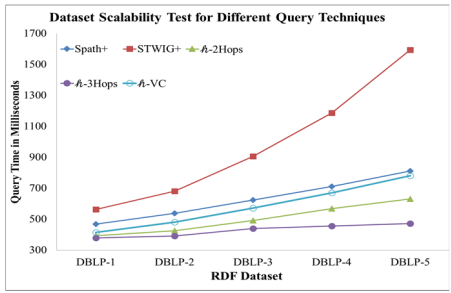
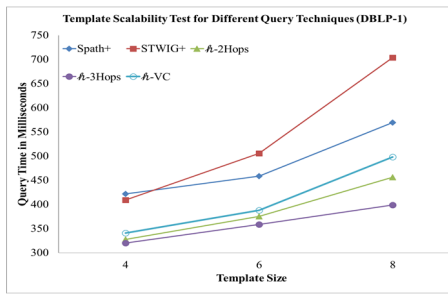

Fig. 5. Dataset Scalability Test    Fig. 6. Template Scalability Test

SP2B and DBLP each has an average node degree of approximately 3, and it is approximately 8 for LUBM and IMDB. Thus, the space increase of NI index by indexing more hops of neighbors is much sharper for LUBM and IMDB compared with SP2B and DBLP.

### 6.2 Query Performance for Different Datasets

For each dataset, 40 random queries with size 6 are generated separately. Query performance results of all query techniques for different datasets are shown in Figure 4. For LUBM, the neighborhood check is not beneficial for most queries due to its coherent graph structure and uniform literal labels. Since the pruning power of the neighborhood check is negligible, the more hops of neighborhood check degrades the performance for LUBM due to its time overhead. The other three RDF datasets, SP2B, DBLP and IMDB have performance patterns similar to each other. As expected, the neighborhood check helps to prune unnecessary intermediate candidates and joins effectively, and, as a result, more hops of neighborhood check results in better performance. Since SPath(NI2) uses the additional neighborhood check, query performance suffers because of it for LUBM, but benefits from it as effective pruning for other datasets. On the opposite, as STWIG+ never uses the additional neighborhood check, it performs well for LUBM, but suffers from the unnecessary intermediate candidates generated for datasets SP2B, DBLP and IMDB. RDF-$\hbar$ algorithm outperforms both SPath(NI2) and STWIG+ since it combines the advantages. Simple queries are processed directly by D-tree candidate generation and joins while complex queries are accelerated by utilizing the additional neighborhood checks. Checking for 2 hop and 3 hop neighbors are more effective in IMDB compared with SP2B and DBLP since 2 hop paths and 3 hop paths are rare which leads to smaller number of query nodes generated with 2 hop or 3 hop neighbors. Compared with $\hbar$-2Hops, $\hbar$-VC works well since the space required is smaller. Performance of $\hbar$-2Hops is slightly worse than $\hbar$-2Hops, as expected.

### 6.3  More Experiments : Scalability Test and Queries with Connection Edges

We perform two scalability tests: dataset scalability and query template scalability. For the dataset scalability, we use DBLP datasets with increasing sizes (from 1 million to 5 million triples) to perform the dataset scalability test with the same set of 40 random queries with size 6. As the dataset contains more triples, partial keywords for literal labels should be re-generated. Results of dataset scalability test are shown in Figure 5. Clearly, STWIG+ suffers the most when the size of the dataset increases since the number of unnecessary candidates and intermediate joins increase dramatically. For randomly generated queries with size 6, 3 hop neighborhood check is sufficient to capture most of the graph structure, and scales best in terms of query performance. $\hbar$-2Hops algorithm has a similar performance pattern with SPath(NI2) since they both use 2 hop neighborhood check. Due to enforcing all queries to utilize the neighborhood check, and not all queries benefit from it, SPath(NI2) has overhead, and ends up performing not as well as the $\hbar$-2Hops algorithm. The $\hbar$-VC algorithm using combined 1 hop and 2 hop neighborhood indexes works well when the RDF dataset is small. But query time increases sharper than SPath(NI2) and $\hbar$-2Hops algorithms. Thus, signature-based pruning is powerful in reducing unnecessary candidates and intermediate joins which scales well with dataset size increases. Here, our dataset scalability test produces different conclusions for the STWIG+ algorithm of [23]. Authors of [23] conclude that the graph size has no significant impact on the response time of the STWIG algorithm when the average node degree is fixed. This difference is due to two factors: (i) the synthetic data generated in [23] has a completely different structure than the DBLP dataset; (ii) STWIG algorithm [23] is a distributed algorithm implemented in a computing cluster rather than a single machine environment.

For the Query Template scalability test, we use random queries with size 4, 6 and 8 (larger template size is rare in real applications). Figure 6 shows performance results for different query techniques for DBLP dataset (due to space limits, the results for SP2B, LUBM and IMDB can be found in [28]). For DBLP, signature-based pruning benefits more as expected when the query template size increases since graph structure becomes more complex and has more query nodes with more hop neighbors. $\hbar$-3Hops algorithm performs the best and $\hbar$-2Hops algorithm works slightly better as compared with the $\hbar$-VC algorithm when the size of the query template increases.

We also performed experiments to test the performance of using different indexes for evaluating queries with connection edges. As expected, using neighborhood indexes with more hops improves the performance of evaluating such queries significantly. The experimental results for the DBLP-1 dataset for queries with connection edges (distance constraint $d_c = 5$) shows that with 3 hops index connectivity check takes 3.6% of the time to evaluate the query, while it takes 41.17%, and 92.45% for 2 hops and 1 hop indexes respectively. More details on these experiments are available in [28].

### 6.4 Choice of Hybrid Algorithm

Choice of neighborhood check for RDF-$\hbar$ algorithm is important for different datasets. For most RDF datasets which can benefit from the neighborhood check, there is a tradeoff between query performance and space efficiency as shown in experiments. The frequency of complex user queries and the frequency of connection edges with large distance constraints have a big impact on the choices of different NI indexes. RDF-$\hbar$ algorithm is more effective when user input queries have different levels of complexity. Query evaluation utilizes the characteristics of the dataset and query templates to choose the appropriate execution plan. Compared with SPath(NI2) and STWIG+, RDF -$\hbar$ algorithm can provide an overall 20% to 30% performance improvements in most cases. The dataset evaluation metrics also give us hints whether a dataset can benefit from signature-based pruning. For a dataset with a high *coherence*, low *specialty* and low *label diversity*, such as LUBM, one should consider $\hbar$-VC first due to little expected benefits from signature-based pruning and best space efficiency.

## 7 Conclusions

In this paper, we evaluate the effectiveness of signature-based pruning on querying graph-structured RDF data using graph templates. Due to the observation that signature-based pruning is not always beneficial, we propose a hybrid algorithm RDF-$\hbar$, which selectively uses neighborhood check based on the characteristics of RDF datasets and query templates. By tuning the parameters, RDF-$\hbar$ algorithm can automatically capture frequent user query patterns and be adjusted to maximize the benefits of signature-based pruning which provide an overall 30% query performance improvement for random generated subgraph queries. Based on the RDF dataset characteristics analysis, we can also identify datasets with respect to the expected level of performance gains from signature-based pruning.

There are several directions for extending this work. Improving the scalability of our framework to even larger datasets is important. This paper's approach scales to RDF datasets with millions of triples which is reasonably large, but there are real datasets with billions of triples. Without proper compression techniques, one can hardly build the NI index on billion node graphs. Some queries can end up with very large number of matches, and a good ranking function is needed to be able to return the most relevant results in rank order.